# Assisted-hydrothermal Synthesis and Characterization of Flower-like ZnO Nanostructures

S. López-Romero[1*], P. Santiago[2], and D. Mendoza[1**]

[1]Instituto de Investigaciones en Materiales, Universidad Nacional Autónoma de México, Departamento de Materia Condensada y Criogenia, Apdo. Postal 70-360, México D.F. 04510 , Coyoacán, México.

[2]Instituto de Física, Universidad Nacional Autónoma de México, Apartado Postal 20-364, 01000 Mexico D.F., Mexico.

(*) sebas@unam.mx

(**) doroteo@unam.mx

**Abstract**

Flower-like nanostructures formed by ZnO nanorods were synthesized and deposited on seeded silicon and glass substrates by a hexamethylenetetramine (HMTA) – assisted hydrothermal method at low temperature (90 $^o$C) with methenamine (($CH_3$)$_6$$N_4$), as surfactant and catalyst. The substrates were seeded with ZnO nanoparticles. The structure and morphology of the nanostructures were studied by means of x-ray diffraction (XRD), high resolution transmission electron microscopy (HRTEM), and scanning electron microscopy (SEM) techniques. Influence of the seed nanoparticle on the formation of the flower-like ZnO nanostructures is demonstrated. The influence of the organic oxygenated chains on the crystalline habit during the growth process is also observed.

Keywords: Zinc oxide; Flower-like nanostructures; Seeded growth; assisted hydrothermal synthesis.


1. Introduction

Zinc oxide is one of the most studied oxide semiconductors and recently has attracted attention due to the possibility of producing several complex nanostructures. These nanostructures might be applied for the fabrication of blue emitting lasers due to their wide band gap (3.37 eV) and high exciton binding energy (60meV), [1] transparent semiconductors, [2] piezoelectric devices, [3] short-wavelength light-emitting devices, [4] blue emitting LEDs, [5] chemical sensors, [6] solar cells, [7] etc. Several methods such as, electrodeposition, [8] evaporation, [9] vapor-liquid-solid (VLS) growth, [10] metalorganic catalyst assisted vapor-phase epitaxy, [11] aqueous thermal decomposition, [12] microwave activated chemical bath deposition (MW-CBD), [13] chemical bath deposition (CBD), [14] hydrothermal-assisted method, [15] etc. have been reported for the production of this kind of nanomaterials. However, hydrothermal-assisted (HTA) method is more convenient over others as it is less expensive with easier composition control, large area deposition, and works at lower temperatures. On the other hand, small changes in any hydrothermal parameters, such as temperature, pH, molar ratio of the precursors, or even reaction time, generate profound influence on the final products .Moreover, using HTA method, ZnO nanostructures of different morphologies could be synthesized. [16] In this work, we report on the production of flower-like nanostructures of ZnO by a two-step surfactant assisted hydrothermal method on different substrates. The nanostructures were deposited on pre-treated silicon and glass substrates by seeding ZnO nanoparticles on their surfaces. It is shown that the pre-treatment of the substrates has a great influence on the growth of the flower-like nanostructures.

2. Experimental

2.1. Materials

The chemical reagents used in this study were analytical reagent grade (Sigma-Aldrich) and used as received without further purification. Silicon wafers (Virginia semiconductor, inc.) with a <100> orientation and glass plates were used as substrates.

2.2 Pre-treatment of the substrates

The process to obtain a colloidal solution to deposit the ZnO nanoparticle seeds on the substrates surface (ZnO seeded substrates) has been described elsewhere [17]. The procedure is based on the sol-gel method. Briefly, zinc acetate [$Zn(CH_3COO)_2$] and cetyltrimethylammonioum hydroxide (CTAOH) were taken as precursor materials. Initially 0.01M of zinc acetate was dissolved in ethyl alcohol and magnetically stirred at 60 °C for 1h. Then, cetyltrimethylammonium hydroxide was mixed into the solution with Zn/CTAOH molar ratio of 1/1.6 and then refluxed at 60 °C for 2h. By direct immersion of silicon and glass substrates into the colloidal solution, the ZnO nanoparticle seeds were deposited onto the substrate surfaces. Subsequently, the substrates were heated in dry air at 300 °C for 12h.

2.3 Preparation of flower-like nanostructures

Zinc nitrate ($Zn(NO_3)_2 \cdot 6H_2O$) was used as precursor and hexamethylenetetramine, also called methenamine (($CH_3)_6N_4$) as surfactant and catalysts. The precursor solution was prepared by dissolving 3.0 g of zinc nitrate and 2.8 g of methenamine in deionized water

under vigorous stirring at 50 °C for 1h to form a 0.01 M equimolar solution. Then, the seeded silicon and glass substrates were immersed in this solution at 90 °C for 2h. It was observed that a white ZnO powder precipitated at the flask bottom. Finally, the substrates were thoroughly washed with deionized water and allowed to dry in air at room temperature.

The reaction mechanisms proposed for the hydrothermal synthesis is already reported by J. Zhung and coworkers [14]. Based on Zhung analysis, the $[Zn(OH)_4]^{2-}$ role is well established and the corresponding chemical reaction for this particular hydrothermal synthesis is as follow:

$$Zn(NO_3)_2 + 2H_2O \xrightarrow{(CH_3)_6N_4,\ 90\ °C} Zn(OH)_2 + 2HNO_3 \qquad (1)$$

$$Zn(OH)_2 \leftrightarrow Zn^{2+} + 2HO^- \qquad (2)$$

$$Zn^{2+} + 2HO^- \leftrightarrow ZnO + H_2O \qquad (3)$$

$$Zn(OH)_2 + 2OH^- \leftrightarrow [Zn(OH)_4]^{2-} \qquad (4)$$

In reaction (1), $Zn^{2+}$ ions are combined with OH⁻ radicals in the aqueous solution to form a $Zn(OH)_2$ colloid through the reaction $Zn^{2+} + 2OH^- \rightarrow Zn(OH)_2$. Later, in the hydrothermal process, the $Zn(OH)_2$ is separated into $Zn^{2+}$ ions and OH⁻ radicals according to reaction (2). Then, ZnO nuclei are formed according to the reaction (3), when the concentration of $Zn^{2+}$

ions and OH⁻ radicals reaches a supersaturation grade. Finally, the growth units of $[Zn(OH)_4]^{2-}$ radicals are obtained through the reaction (4). The dissolution-nucleation cycle according to reactions (5) and (6), respectively produces:

$$[Zn(OH)_4]^{2-} \leftrightarrow Zn^{2-} + 4OH^- \qquad (5)$$

$$Zn^{2+} + 2OH \leftrightarrow ZnO + H_2O \qquad (6)$$

3. Characterization

Morphology of the sample was studied using a JEM5600-LV scanning electron microscope. The single-crystal structure of the ZnO nanorods was studied using a JEOL FEG 2010 Fast TEM electron microscope with a 2.1Å resolution (point to point). The x-ray diffraction (XRD) pattern of the ZnO nanorods was obtained with a x-ray diffractometer (SIEMENS D 5000) using the $CuK_α$ (1.5406Å) radiation, with a scanning speed of 1° per min at 35 KV and 30 mA.

4. Results and discussion

Figure 1 shows the XRD patterns of the flower-like ZnO nanostructures prepared by the HMTA-assisted hydrothermal method, grown on silicon (Fig. 1a) and glass (Fig. 1b) (pre-treated) substrates along with the white precipitated powder at the bottom of the reaction flask (Fig. 1c), respectively. It is clearly observed that the XRD patterns are similar for the three samples; therefore, there is no evidence of variation in the relative orientations of the material. The XRD patterns of the samples grown on silicon (Fig. 1a), glass (Fig. 1b) and precipitated powder confirm that all of them correspond to ZnO with Wurtzite structure. All diffraction peaks of the three samples were indexed to the hexagonal phase of ZnO having lattice parameters a = 3.249 and c = 5.206 Å (JCPD file No 36-1451). The XRD results indicate that the nature of the substrates have no effect on the crystalline structure of the deposited material.

The SEM micrograph of the glass substrate after immersion into the colloidal solution is shown in figure 2a. The panoramic view in figure 2a reveals large agglomerations of ZnO seeds while there are other zones with almost no material. Such inhomogeneities in seed distribution are reflected in the growth of flower-like structures on the seeded substrates (figures 2b and 2c).

Figures 2b and 2c show the typical low magnification SEM images of the flower-like ZnO nanostructures deposited on seeded silicon and glass substrates, respectively. They consisted of nanorods emerging from a common point in all directions (inset of Fig. 2b). The nanorod sizes are between 200-300 nm in diameter, and 1-2 µm in length. Figure 2d is a SEM image of the precipitated white material. It is clearly seen that the ZnO nanorods exhibit hexagonal habit of the Wurtzite phase. The powder sample reveals nanorods wth morphology corresponds to a continuous bush-like precipitate without the presence of flower-like nanostructures. This result indicates that the ZnO nanoparticle seeds act as precursor for the growth of flower-like ZnO nanostructures as we see in the substrates.

In a closer view for the silicon supported sample, it is possible to observe different ZnO rods morphologies. In figure 3a, for example, flower-like morphologies coexist with planar belt shapes. The SEM micrograph 3b shows some ZnO rods with evident hexagonal habit (pointed with white arrows). However, in figure 4a it is possible to detect structures with planar morphologies, where the angle between the rods is 90° forming an "L" configuration. These types of structures are not possible to grow with hexagonal habit. Therefore, it is likely that the high organic concentration inhibit the specific grown planes in the ZnO nanorods generating those morphologies. This confirms the role of acetate in the growth process, where the crystallographic independent Zn(II) atoms take on different coordination morphologies. [18,19] Therefore, it is possible to observe another minority phase exhibiting very different morphological characteristics as in figure 3 and 4. The coordination environments of metal atoms induce different coordination structures. Therefore, each Zn(I) atom adopts a distorted octahedral geometry when it is coordinated by four carboxylate oxygen atoms, while Z(II) atom generates a distorted trigonal geometry when it is ligated to three different carboxylate oxygen atoms [17,18]. Though Ovalle and coworkers [19] have reported a distorted octahedral geometry of ZnO, the role of the acetate on the morphology is still under study. However, It is well known that $Zn^{+2}$ can easily coordinate with oxygenated species affording organometallic compounds with different geometries around the metal center. [18,19] With this in mind, we can propose that the high contents of organic oxygenated chains can modify the crystalline habit during the growth process. This can explain the high angle between the ZnO rods. On the other hand, the absence of organometallic compound in the growth process of flower-like morphologies confirms the acetate role in its early stage. From the micrographs it is possible to observe that there is a predominant Wurtzite phase with a characteristic hexagonal habit as we expected in the flower-like structures.

Figures 5a and 5b are the electron diffraction pattern and HRTEM image of a nanorods belonging to the flower-like structures, respectively. These results indicate a good crystalline quality of the obtained material, which is consistent with the x-ray results shown in figure 1. The HRTEM image reveals that the interplanar spacing in the crystalline nanorods is 0.26 nm, which corresponds to the distance between two (002) planes of the hexagonal ZnO phase, indicating the preferential growth along the [002] direction (c-axes). A probable explanation for the role played by the seed ZnO nanoparticles on the nucleation and growth of the flower-like nanostructures has been suggested by H. Zhou et al. [20] Thus, the ZnO nanoparticles promote the nucleation process of ZnO nanorods in the flower-like structures. Moreover, the as-prepared ZnO nanoparticles serve as nuclei since they have the same crystalline structure and similar lattice parameters as of the ZnO nanorods.

5. Conclusions

In this work flower-like ZnO nanostructures were synthesized by the HMTA-assisted hydrothermal method on seeded silicon and glass substrates. The flower-like nanostructures consist of nanorods emerging from a common point. The ZnO nanorods are highly crystalline and of hexagonal wurtzite phase. The pre-treatment of the substrates by adding ZnO nanoparticles is a facilitating step to obtain flower-like ZnO nanostructures. $Zn^{+2}$ ions can easily coordinate with oxygenated species affording organometallic compounds with different geometries around the metal center, generating ZnO nanostructures of different morphologies. However, the process needs further investigation to be confirmed.

Acknowledgements
The authors wish to thank L. Baños (IIM) for his support in carrying out x-ray study, Carlos Flores (IIM) and Luis Rendón (IF) for their support in the electron microscopy characterization. The authors also thank to M. Sc. Pilar Ovalle for helpful discussions. The authors are also thankful to the Central Microscopy facilities of the Institute of Physics,

UNAM, for providing the microscope tools used in this work. This work was financially supported by CONACyT-Mexico (Grant number 52715) and CUDI-CONACyT grant.

**Figure captions.**

Fig. 1. x-ray diffraction pattern of the flower-like ZnO structures grown on silicon substrate (a), on glass substrate (b), and of the precipitated white powder (c). The diffraction patterns in all cases show Wurtzite crystalline structure.

Fig. 2. SEM micrographs of (a) ZnO seeds on glass substrate, (b) ZnO flower-like structures on silicon. At the inset is possible to observe an enlargement of a flower-like structure, and (c) on glass substrates. (d) SEM image of the precipitated powder.

Figure 3. SEM micrographs of a) ZnO structures in an early stage of the grown process. b) Flower-like structures with hexagonal habit coexist with planar and flatted morphologies.

Figure 4. a) SEM micrographs of cross and L ZnO rods configuration. b) Flower-like structures with hexagonal habit and flatted rods. c) "L" configuration which confirms the acetate role on the grown process. d) Cross configuration arrangement of ZnO rods.

Fig. 5 (a) Electron diffraction pattern from a single nanorod. (b) HRTEM image of a nanorod belonging to the flowerlike nanostructure. The nanorod grows along [002] direction.

Figure 1. S. López-Romero, P. Santiago, D. Mendoza. Assisted-hydrothermal synthesis and characterization of ZnO flower-like nanostructures.

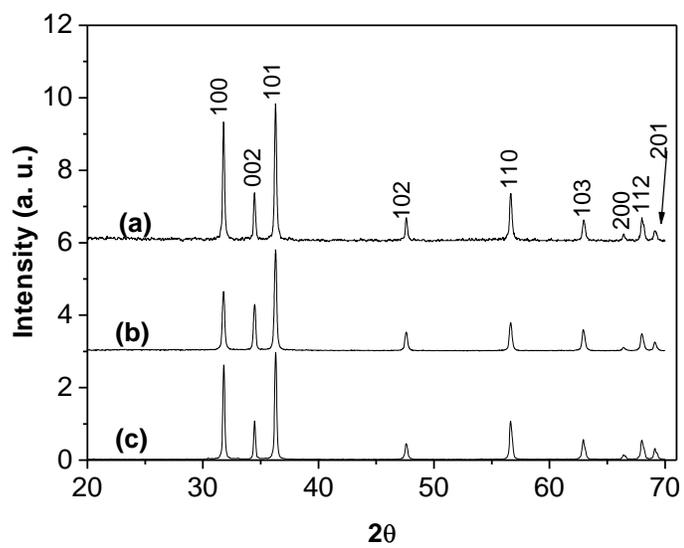

Figure 2. S. López-Romero, P. Santiago, D. Mendoza. Assisted-hydrothermal synthesis and characterization of ZnO flower-like nanostructures.

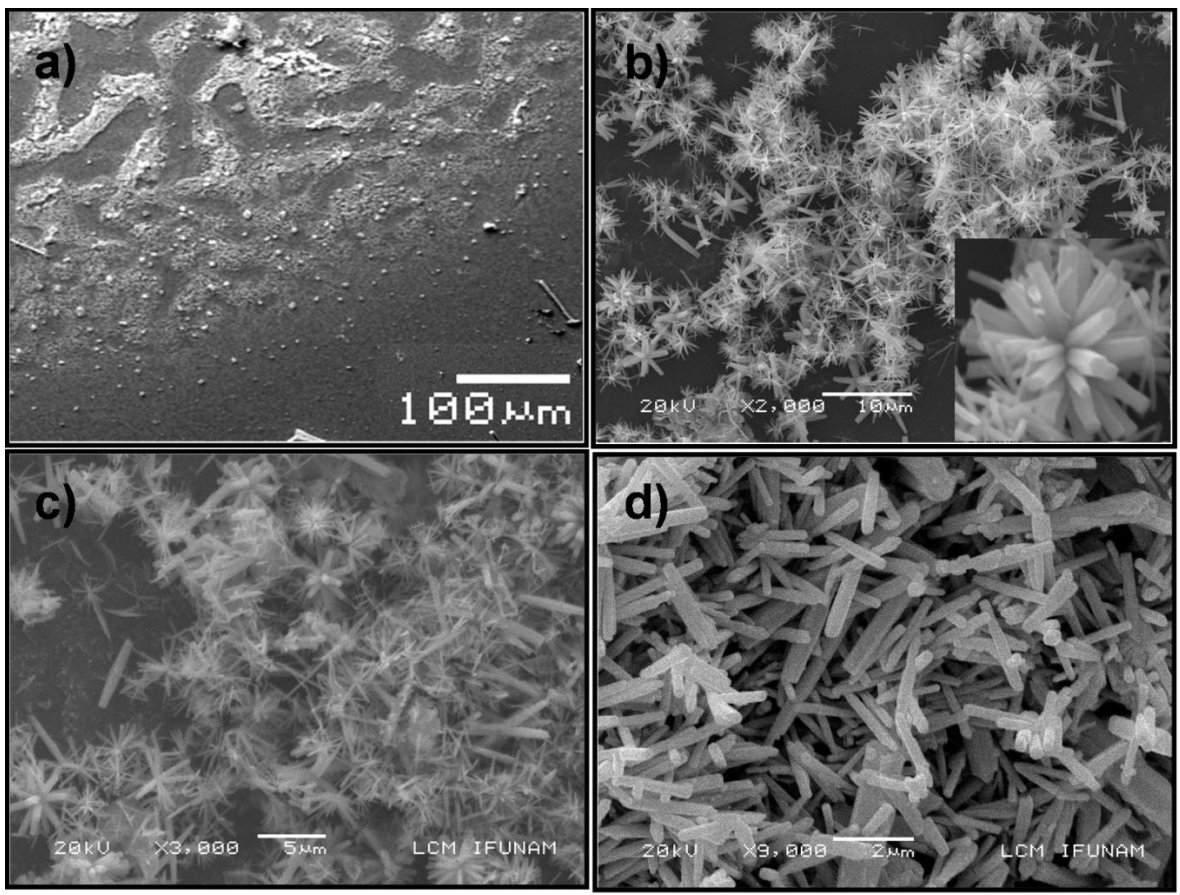

Figure 3. S. López-Romero, P. Santiago, D. Mendoza. Assisted-hydrothermal synthesis and characterization of ZnO flower-like nanostructures.

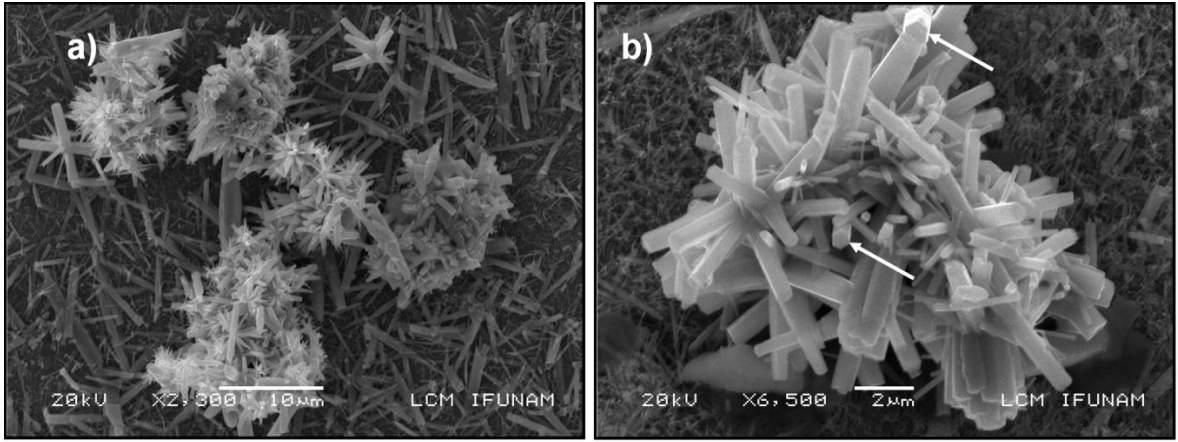

Figure 4. S. López-Romero, P. Santiago, D. Mendoza. Assisted-hydrothermal synthesis and characterization of ZnO flower-like nanostructures.

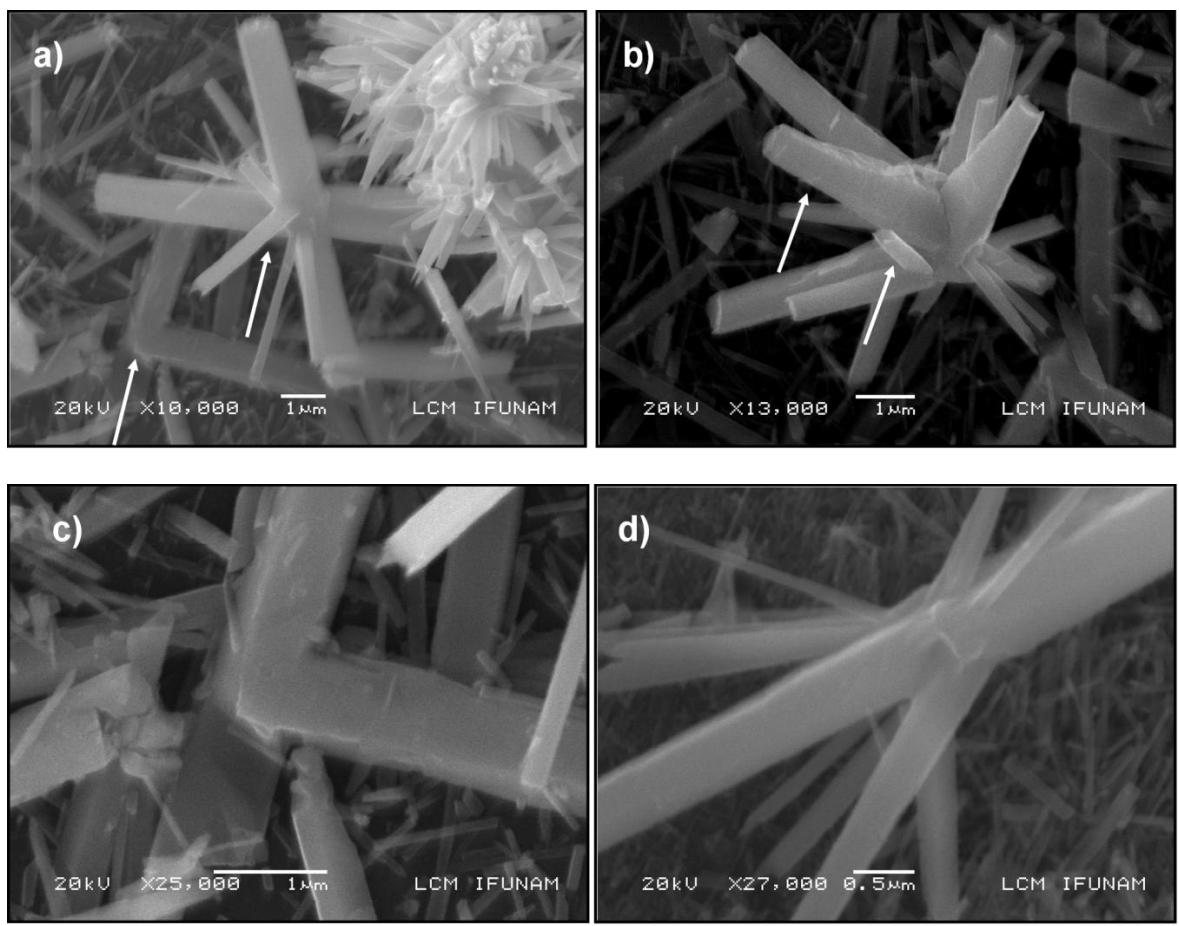

Figure 5. S. López-Romero, P. Santiago, D. Mendoza. Assisted-hydrothermal synthesis and characterization of ZnO flower-like nanostructures.

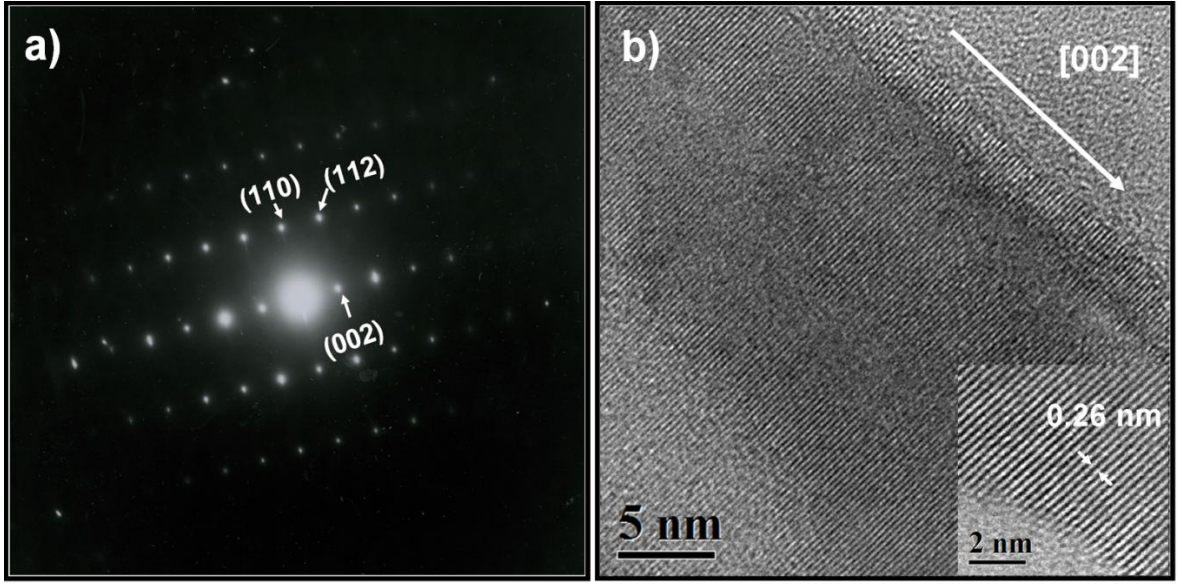